# Three axis vector atomic magnetometer utilizing polarimetric technique

Swarupananda Pradhan

*Laser and Plasma Technology Division, Bhabha Atomic Research Centre, Mumbai-85, India*
*Homi Bhabha National Institute, Department of Atomic Energy, Mumbai-94, India*

*Corresponding addresses: spradhan@barc.gov.in, pradhans75@gmail.com*

The three axis magnetic field measurement based on the interaction of a single elliptically polarized light beam with an atomic system is described. The magnetic field direction dependent atomic responses are extracted by the polarimetric detection in combination with laser frequency modulation and magnetic field modulation techniques. The magnetometer offers additional critical requirements like compact size and large dynamic range for space application. Further, the three axis magnetic field is measured using only reflected signal from the polarimeter, thus can be easily expanded to make spatial array of detectors or / and high sensitivity field gradient measurement as required for biomedical application.

The highly sensitive atomic magnetometers have drawn much attention due to their potential uses in wide range of field encompassing civilian to defence to biomedical application. The recent progresses in these contemporary devices also indicate that they have superior qualities compared to state-of-art SQUID magnetometer in terms of functionality as well as flexibility [1, 2]. Though in a developing stage, there is no doubt that these atomic magnetometers will be imperative for high sensitive magnetic field measurement for a prolonged time in near future. There are several competing schemes for operation of the atomic device and the supremacy of a scheme can be adjudged by salient features such as high sensitivity, possible compact operation, large dynamic range, capability for three axis field measurement and possible expansion to arrays of detectors. The last attribute is very important for viable biomedical application, particularly in magneto-encephalography.

The various atomic magnetometers has their own advantages, however majority of them rely on the zero field resonance observed in Hanle type experimental procedure [3, 4]. The origin of these resonances has been extensively studied. Depending on the experimental parameters, the atomic coherence effect and/or optical pumping followed by ground state redistribution of population are described as the responsible mechanism behind the observed resonances [5-7]. In general, the quantum interference effect has been established as the governing mechanism for the observed dark and bright resonance for $F_g \to F_e \leq F_g$ and $F_g \to F_e > F_g$ transition respectively, while working with linearly polarized light.

However, in presence of a transverse magnetic field (T-field), the optical pumping and subsequent redistribution of Zeeman state population dominates over the quantum interference, and a decrease in the transmitted light intensity is invariably observed in a buffer gas environment. Recently, we have utilized this phenomenon in combination with frequency modulation (FM) spectroscopy for operation of an atomic magnetometer [8]. Here, the transmitted FM signal has been used for frequency stabilization, where as the reflected FM signal provides magnetic field measurement. The operation of this magnetometer with large dynamic range has been demonstrated by feed back control of the bias magnetic field. Also, it has been pointed out that the magnetic field modulation (MM) signal can be equally used for magnetometry due to comparable signal to noise ratio (SNR).

The motivation behind this work is to establish credible method for measurement of three axis magnetic field [9, 10], while retaining the advantages of autonomous frequency stabilization and large dynamic range [8]. Another important aspect is to have provision for expansion of this device to arrays of detectors for spatially resolved or/and gradient magnetic field measurement.

Though T-field has been used for enhancement of the zero field signal amplitude, it is not an encouraging option for three axis magnetic field measurement. A linearly or elliptically polarised light interacting with an atomic ensemble near zero field is the simplest and idealistic choice, that essentially breaks the symmetry of the system for possible three axis magnetic field measurement. In generally, we always get enhanced transmitted and reflected zero field resonance signal in a polarimetric detection system (see set-up description) as compared to direct signal (without detection PBS) irrespective of transition, ellipticity or presence of T-field. As the elliptically polarized light induces strong optical activity in the system, the reflected light is gainfully enhanced. Further the reflected and transmitted light through the polarimeter near two-photon resonance has distinct features that bring additional flexibilities as well as newer possibilities [8, 11-13]. In view of these above considerations, we have chosen an elliptically polarized light resonantly interacting with an atomic system near zero-field for measurement of three axis magnetic field.

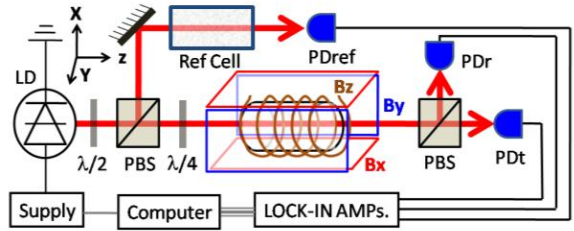

**Figure-1:** Top view of the schematic set up for three axis magnetic field measurements is shown here. The polarizer PBS before the cell and the detection PBS are kept parallel with each other along the x-axis (see text for more details).

The schematic experimental diagram is shown in figure-1. We have preferred to perform the experiment on a table top set-up rather than the compact system described in ref-8. The experimental geometry is slightly changed, while most of the experimental parameters are similar. Briefly, the polarizer PBS and the detection PBS axis are kept parallel to each other along the x-axis in horizontal plane. The laser propagation and vertical directions are along z and y axis respectively. A modulation is applied to the injection current of the VCSEL laser operating near 795 nm. The reflected and transmitted signals across the detection PBS are phase sensitively detected with respected to this modulation. These signals are termed as reflected and transmitted FM signal respectively. The quarter wave plate axis is kept at $20^0$ with respect to x-axis to generate an elliptically polarized light. Due to large ellipticity as compared to ref-1, the transmitted FM signal has little dependence on the magnetic field and is neutralized by subtracting a fraction of reflected FM signal from it [8, 13]. The laser is either locked to this generated signal or to the FM signal from the reference cell without any difference. Thus for the operation of the magnetometer, the reference set-up shown in figure-1 is not required. The laser is frequency stabilized in between Rb-85 F=3 → 3 and F=2→2 transition (~1.3 GHz away from each transition). The temperature of the atomic cell is maintained at $48^0$ ($\pm.5^0$) C.

The atomic cell containing Rubidium atoms in natural isotopic composition and 25 Torr $N_2$ buffer gas is kept in there axis magnetic coils as shown in figure-1. The atomic cell and the three axis magnetic coils are enclosed inside several layers of mu metal sheet. The bias currents through the magnetic coils are controlled using a computer program and added with modulations [9, 14, 15]. The modulation frequency to Bx, By, and Bz fields are 69 Hz, 79 Hz and 55 Hz respectively. The modulation amplitude to each of the coils is ~ 90 nT. The reflected and transmitted photodiode signals are phase sensitively detected with respect to the magnetic field modulation applied along u (x, y, z) direction and are represented as MMuR and MMuT respectively. All the six MMu signals are simultaneously acquired.

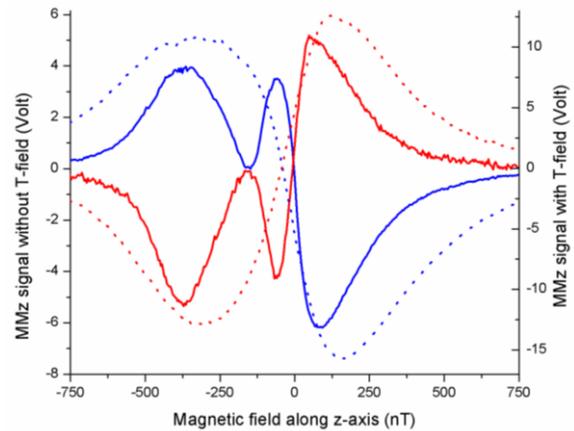

**Figure-2:** The reflected (blue) and transmitted (red) magnetic field modulation signal in presence (dotted line) and absence (solid line) of transverse field (T-field) are shown here. Though signal amplitude is decreased in absence of T-field, the slope of the signal has improved significantly.

As has been pointed out in ref-8, either reflected FM or magnetic field modulation (MM) signal in presence of a T-field can be used for magnetic field measurement. However in absence of T-field, both of these signal has reduced amplitude, but the MM signal has better SNR. Thus MM signal is a better choice for magnetometry in absence of T-field as compared to FM signal. The MMzR and MMzT signal with and without T-field are shown in figure-2. Though the signal amplitude is reduced in absence of T-field, the slope of the signal has significantly improved. Thus sensitivity of magnetometer using MMz signal is expected to remain comparable for with and without T-field.

Figure-2 illustrates a kink in the profile of MMz signal in absence of T-field. Since we are measuring the slope of the actual signal profile, the observed kink is due to change in the slope and reflect asymmetry in the signal profile. Such asymmetry in the zero field resonance has been observed and studied earlier [5, 16, 17]. Due to slope measurement, the asymmetry in the line profile has prominently appeared in figure-2. However, we are interested in small range of magnetic field (±50 nT) where we get a smooth slope for the MMz signal.

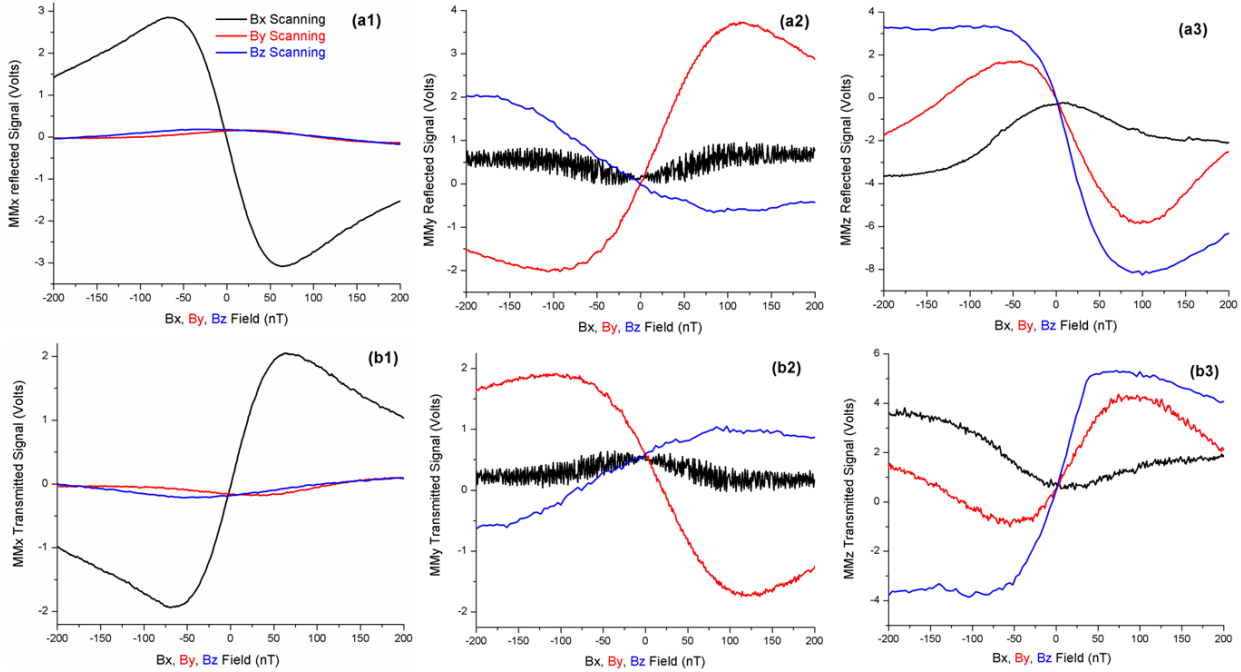

**Figure-3:** The reflected (a1,a2,a3) and the transmitted (b1,b2,b3) MMu (u = x,y,z) signal for the magnetic field scanned through zero-field along x-axis (black), y-axis (red) and z-axis (blue) are shown in absence of T-field. It is interesting to note that the amplitude of the MMu signal is invariably maximized for scan and modulation in the same direction as compared to orthogonal direction.

In order to have a reliable three axis magnetic field measurement, the atomic system should have the desired asymmetry to respond differently with respect to the direction of the magnetic field. The experimental regime for three axis magnetic field measurement is described for and successfully implemented to SERF magnetometer [9]. In our experimental geometry, the major axis, minor axis and the laser propagation direction introduced the desired asymmetry. Further we work in a regime that satisfies the criteria given in ref-9. The responses of the atomic system along three orthogonal directions are acquired through the MMu signal as shown in figure-3. The magnetic field is scanned in one direction while keeping field along orthogonal direction near zero. The oscillation seen in the reflected and transmitted MMy signal for Bx scanning is due to close modulation frequency applied to Bx and By field. It disappears as the Bx value is brought near zero and can be suitably exploited for magnetometery. More importantly, it is interesting to note that both reflected and transmitted FM signal has highest amplitude for modulation and the scan field in the same direction as compared to in orthogonal direction. This feature is sufficient for three axis magnetic field measurement and can be realized as follows.

The magnetic field modulation signal in any direction (u) has three contributions to it.

$$MMu = \left[\frac{\partial MMu}{\partial Bx}\right]Bx + \left[\frac{\partial MMu}{\partial By}\right]By + \left[\frac{\partial MMu}{\partial Bz}\right]Bz$$

For three axis magnetic field measurement one has to iteratively control the Bx, By and Bz field to bring the MMu signal to zero. This can be carried out through active software or electronic servo controller. Once all the MMu signals are brought to zero, the amplitude and the polarity of the current flowing through the magnetic coils are measure of the magnetic field amplitude and polarity in the respective direction. The condition for convergence of iteration can be easily found out to be $\frac{\partial MMu}{\partial Bu} > \frac{\partial MMu}{\partial Bu'}$, where $u, u' = $ x, y, z and $u \neq u'$. These slopes of the MMu signals are measured by changing the magnetic field in any direction by ±25 nT, while keeping other orthogonal field to zero value and subsequently repeating for other directions. The slope of the reflected MMu (MMuR) and transmitted MMu (MMuT) signal for change in magnetic field in three orthogonal directions are tabulated in table-1. It may be seen from the data that both reflected and transmitted MMu signal satisfies the condition for convergence of iteration comfortably. We have manually found that within 2-3 iteration, it is possible to bring the three axis magnetic field to zero value. Further the maximum value of slopes is in the same order and in combination with SNR reflects the field sensitivity in that direction.

| Bu | $\frac{\partial MMxR}{\partial Bu}$ (mV/nT) | $\frac{\partial MMxT}{\partial Bu}$ (mV/nT) | $\frac{\partial MMyR}{\partial Bu}$ (mV/nT) | $\frac{\partial MMyT}{\partial Bu}$ (mV/nT) | $\frac{\partial MMzR}{\partial Bu}$ (mV/nT) | $\frac{\partial MMzT}{\partial Bu}$ (mV/nT) |
|---|---|---|---|---|---|---|
| Bx | **86.8** | 56.5 | **3.3** | 2.1 | **13.5** | 10.1 |
| By | **2.7** | 1.7 | **53.1** | 32.3 | **90.2** | 64.8 |
| Bz | **2.3** | 3.1 | **12.4** | 7.1 | **146.2** | 115.5 |

**Table-1:** Rate of change of MMu (u = x, y, z) reflection and transmission signal with respect to change in the magnetic field along three orthogonal directions.

An important feature of this magnetometry geometry is that the three axis magnetic field measurement can be performed using either reflected or transmitted MMu signal. Use of only reflected MMu signal for three axis magnetic fields has a very unique advantage as the linearly polarized transmitted signal can be passed through another set of quarter wave plate, atomic cell with three axis field control and detection PBS for measure magnetic field at a spatially different location. In principle, it can be expanded to an array of detectors. This is an essential requirement for viable magneto-encephalography, where spatially resolved magnetic field measurement is required. In such a configuration, the sensitivity can be further improved by gradient field measurement. Since the laser frequency stabilization is performed with the transmitted FM signal, a part of the light after the first detection PBS or transmitted light by the last detection PBS can be used for this purpose.

The established atomic magnetometer has overcome an important limitation of three axis vector operation as desired for space and other related applications. Since this atomic magnetometer operates with a single elliptically polarized beam and laser frequency can be stabilized in situ without requirement of additional set-up, it can be made compact in size. The feedback control can be used for realizing large dynamic range as demonstrated in ref-8. Another important aspect is the possible expansion to array of detectors and / or gradient field measurement, as only reflected signal by detection PBS is sufficient for three axis magnetic field measurements.